\documentclass[letter,twocolumn]{revtex4-1}
\usepackage[dvipdfmx]{graphicx}
\usepackage{braket}

\usepackage{amsmath,amssymb}
\usepackage{color}

\usepackage{bm}
\usepackage{ulem}

  \newcommand{\average}[1]{\ensuremath{\langle#1\rangle} }

\begin{document}

\title{Acoustic Rashba--Edelstein effect}
\author{Takumi Funato}
\affiliation{Center for Spintronics Research Network, Keio University, Yokohama 223-8522, Japan}
\affiliation{Kavli Institute for Theoretical Sciences, University of Chinese Academy of Sciences, Beijing, 100190, China}
\author{Mamoru Matsuo}
\affiliation{Kavli Institute for Theoretical Sciences, University of Chinese Academy of Sciences, Beijing, 100190, China}
\affiliation{CAS Center for Excellence in Topological Quantum Computation, University of Chinese Academy of Sciences, Beijing 100190, China}
\affiliation{RIKEN Center for Emergent Matter Science (CEMS), Wako, Saitama 351-0198, Japan}
\affiliation{Advanced Science Research Center, Japan Atomic Energy Agency, Tokai, 319-1195, Japan}
\date{\today} 
\begin{abstract}
We theoretically study the mechanical induction of the spin density via the Rashba spin--orbit interaction (SOI). 
The spin density in the linear response to lattice distortion dynamics is calculated based on the microscopic theory. 
We reveal that there are two mechanisms of spin induction: one is the acoustic Edelstein effect (AEE) from the acceleration of the lattice dynamics and the other is caused by the Rashba spin--vorticity coupling (RSVC). 
We find that the AEE induces a more efficient spin-to-charge conversion in comparison with the conventional electric Edelstein effect.
The induced spin density due to the RSVC is expressed as a Berry curvature-like quantity; therefore, it can be attributed to the spatial symmetry breaking due to the Rashba SOI.
Our work demonstrates high-efficiency spin generation in Rashba systems. 
\end{abstract}

\maketitle

The spin-orbit interaction (SOI), which represents coupling of orbital motion and spin angular momentum, plays a significant role in the interconversion between spin and charge in the field of spintronics. 
In particular, the Rashba SOI, which arises from spatial inversion symmetry breaking, is expected to provide a highly efficient spin--charge conversion.
In Rashba systems, the spin degeneracy of electronic energy bands is split, and the electron momentum direction and spin direction are fixed in each band, a phenomenon known as the spin--momentum locking\cite{bychkov1984,rashba1,rashba2,ast2007,landolt2012,eremeev2012,review2004,manchon2015}.
When an external electric field is applied, the energy bands shift, and a non-equilibrium spin density is induced, which is known asthe Edelstein effect (EE)\cite{lyanda-geller1989, pikus1989,edelstein1990,sanchez2013,fan2014,nomura2015,nakayama2016,isasa2016,karube2016}.
The direct/inverse EE provides high efficient interconversion between charge and spin and is expected to substitute the direct/inverse spin Hall effect\cite{d'yakonov1971,hirsch1999,zhang2000,rashba2003,dimitrova2005,shi2006,saitoh2006,sinova2015}.
Investigation of the Rashba SOI is essential for the realization of efficient and versatile spintronics devices.

Exploring different methods to generate a spin current is also an important topic in the field of spintronics.
Recently, the generation of spin currents through the conversion from mechanical angular momentum to spin angular momentum has attracted considerable attention.
As one of its underlying mechanisms, the spin--vorticity coupling (SVC), the coupling between the electron spin and the effective magnetic field associated with the rigid rotation, has been proposed\cite{barnett,ed,matsuo1,matsuo2,matsuo3,matsuo4}.
Spin-current generation via the SVC has been observed in a shear flow of liquid metals (spin hydrodynamic generation)\cite{svc_expt1,svc_expt2,svc_expt3,svc_expt4,svc_theo1} 
and surface acoustic waves\cite{kobayashi2017,kurimune1,kurimune2,tateno2020,tateno2}.
Alternatively, mechanical spin-current generation via the SOI has been investigated both theoretically and experimentally\cite{funato2018,helicity,kawada2021}.
It is suggested that the SOI mechanism can be dominant in strong SOI systems, and it provides spin currens with a characteristic symmetry.
Spin currents are expected to be generated efficiently via mechanical means in Rashba systems;
however, the conversion of mechanical rotation into spin via the Rashba SOI still remains an open problem.

In this work, we theoretically studied the induction of spin density from lattice distortion dynamics in two dimensional (2D) and three-dimensional (3D)  Rashba electron systems.
To treat a couple of lattice distortion dynamics and the electrons, we performed the local coordinate transformation and calculated the spin density in linear response to the lattice displacement.
The results indicate that there are two mechanisms through which the spin density induction occurs: one is the acoustic Edelstein effect (AEE) due to the lattice acceleration and the other is the Rashba spin--vorticity coupling (RSVC) due to the lattice vorticity.
The AEE was found to provide a more efficient charge-to-spin conversion than the conventional EE.
The spin density due to the RSVC includes terms that are independent of the magnitude of the Rashba SOI.
This spin density can be attributed to the spatial symmetry breaking due to the Rashba SOI and is expressed in terms of the Berry curvature-like quantity.
It is suggested that the spin density induced by the RSVC is comparable to or larger than that induced by the conventional SVC.
Therefore, the combination of the RSVC and the conventional SVC provides a more efficient spin generation in Rashba systems.

We model 2D and 3D electrons scattered by impurities in the crystal according to the following equation:
\begin{align}
\hat H= \int d^3r'& \hat{\psi} ^{\prime \dagger}(\bm r',t)\biggl[
\frac{p^2}{2m} +  V (\bm r') \nonumber \\ 
&+ \lambda _{\text{so}} \bm \sigma \cdot [\nabla ^{\prime}V_{\text{p}}(\bm r')\times \bm p
\biggr] \hat{\psi} ^{\prime}(\bm r',t)
,
\end{align}
where $\hat \psi^{\prime \dagger}$ and $\hat \psi^{\prime}$ are the electron field operators, $V = V_{\text{p}}  + V_{\text i} $ is the total potential which combines the lattice periodic potential $V_{\text p}$ and the impurity potential $V_{\text i}$,
$\lambda _{\text{so}}$ is the SOI strength and $\bm \sigma =(\sigma ^x,\sigma ^y,\sigma ^z)$ are the Pauli matrices.
Assuming that an acoustic wave is applied, the Hamiltonian is then modulated according to the following equation:
\begin{align}
\hat H_{\text{lab}}= &\int d^3r' \hat{\psi} ^{\prime \dagger}(\bm r',t)\biggl[
\frac{p^2}{2m} +  V (\bm r'-\delta \bm r) \nonumber \\ 
&+ \lambda _{\text{so}} \bm \sigma \cdot [\nabla ^{\prime}V_{\text{p}}(\bm r'-\delta \bm r)\times \bm p
\biggr] \hat{\psi} ^{\prime}(\bm r',t)
,
\end{align}
where  $\delta \bm r(\bm r',t)$ is the lattice displacement vector.
The effects of the lattice distortion are taken into account by performing a local coordinate transformation from the laboratory frame to the ``material'' frame, i.e. $\bm r=\bm r'-\delta \bm r(\bm r',t)$; the ``material frame" is locally fixed to the moving lattice\cite{funato2018,helicity,tsuneto}.
Specifically, we consider the connection $\nabla _i^{\prime}=(\partial r_i/\partial r'_j)\nabla _j$ and the renormalization of the wave functions, i.e. $\hat \psi ^{\prime}(\bm r',t)=\left| \partial \bm r'/\partial \bm r \right| \hat \psi (\bm r,t)$, where $\left| \partial \bm r'/\partial \bm r \right|$ is the Jacobian and $\hat \psi $ is the wave function of the material frame.
The Lagrangian is given by the following equation:
\begin{align}
    \hat L 
    &=\int  \psi ^{\dagger}(\bm r,t)i\partial _t\psi(\bm r,t)d^3r - \hat H_{\text{mat}},
\end{align}
where $\hat H_{\text{mat}}$ is the Hamiltonian given in the material frame.

Assuming a free-electron system with spatial symmetry breaking along $z$-direction, the second quantized total Hamiltonian up to the first order in $\delta \bm r$ is given by 
$\hat H_{\text{mat}} = \hat H+\hat H'(t)$,
where the first term is the time-independent Hamiltonian, and the second term is the perturbation Hamiltonian due to the acoustic waves.
The time-independent term $\hat H=\hat H_0 + \hat H_{\text i}$ consists of the unperturbed term (kinetic energy and Rashba SOI), $\hat H_0=\sum_{\bm k} c_{\bm k}^{\dagger}\left[k^2/2m+\alpha _{\text R}\hat z \cdot (\bm k\times \bm \sigma )\right] c_{\bm k}$, and the impurity potential term, $\hat H_{\text i}=\sum_{\bm k,\bm k'} V_{\bm k'-\bm k}c_{\bm k'}^{\dagger}c_{\bm k}$,
where $c^{\dagger}_{\bm k}(c_{\bm k})$ is the electron creation (annihilation) operator, $\alpha _{\text R}$ is the strength of the Rashba SOI, and $V_{\bm k}$ is the Fourier component of the impurity potential $V_{\text i}(\bm r)$.
The perturbation Hamiltonian consists of two parts, namely, $H'(t)=H'_{\text k}(t)+H'_{\text r}(t)$, where $H'_{\text k}$ is the modulation of the kinetic energy and the time derivative\cite{tsuneto,funato2018}, and $H'_{\text r}$ is the modulation of the Rashba SOI\cite{helicity}:
\begin{align}
\hat H'_{\text k}(\omega ) &= -\sum_{\bm k,\bm q} c^{\dagger}_{\bm k_+} w^{\text k}_jc_{\bm k_-} u^j_{\bm q,\omega},\\
\hat H'_{\text r}(\omega ) &= -\sum_{\bm k,\bm q} c^{\dagger}_{\bm k_+} w^{\text r}_jc_{\bm k_-} u^j_{\bm q,\omega},
\end{align}
where $\bm k_{\pm}=\bm k\pm \frac{\bm q}{2}$ and $\bm u _{\bm q,\omega}$ are the Fourier components of the lattice velocity, $\bm u(\bm r,t)=\partial _t \delta \bm r(\bm r,t)$.
$w^{\text k}_j$ and $w^{\text r}_j$ are the vertex parts of the perturbation Hamiltonian:
\begin{gather}
w^{\text k}_j =\left( 1-\frac{\bm v\cdot \bm q}{\omega}\right) k_j, \\
w^{\text r}_j = \frac{\alpha _{\text R}}{\omega}[\bm k\cdot (\bm \sigma \times \bm q)\delta _{jz} - \hat z\cdot (\bm \sigma \times \bm q) k_j].\label{wr}
\end{gather}
Throughout this study, we assume that the wavenumber $\bm q$ and the frequency $\omega$ are much smaller than the inverse of the mean-free path $l$ and the relaxation time of the electrons, i.e. $q\ll l^{-1}$ and $\omega \ll \tau ^{-1}$, respectively.

Assuming a uniform random distribution of the impurities and a delta-function like an impurity potential, $V_{\text i}(\bm r)=u_{\text i}\sum_j\delta (\bm r-\bm R_j)$, where $\bm R_j$ is the position of the $j$-th impurity,
the impurity averaged equilibrium retarded/advanced Green functions with impurity scattering included by the Born approximation are given by the following equation:
\begin{align}
    G^{\text{R/A}}_{\bm k}(\epsilon) = \frac{1}{2}\sum_{s =\pm}\frac{1+s \Gamma _{\bm k} }{\mu+\epsilon -\epsilon_{s}\pm i\gamma},
\end{align}
where $\Gamma _{\bm k}=\epsilon _{z\beta l} \sigma ^{\beta}k_l/k_{\text t}$ with the Levi-Civita symbol $\epsilon _{z\beta l}$, $\epsilon _{s}=\frac{k^2}{2m}+s\alpha_{\text R}k_{\text t}$ is the eigenenergy with $k_{\text t}=(k_x^2+k_y^2)^{1/2}$ , $\mu$ is the chemical potential, $\gamma = \frac{\pi}{2}n_{\text i}u_{\text i}^2\nu$ is the damping rate with the impurity concentration $n_{\text i}$ and the Fermi-level density of state $\nu=2a_1\nu _0$.
The parameter $a_1$ and the density of states for free-electrons $\nu _0$ are defined below Eq.~(\ref{she1}).
Here, we use the ensemble average for impurity positions $\average{V_{\bm k}V_{\bm k'}}_{\text{av}}=n_{\text i}u_{\text i}^2\delta _{\bm k,\bm k'}$.

The spin density and spin-current density operators are given by the following equation:
\begin{gather}
    \hat j^{\alpha}_{\text s,0}(\bm q) = \hat \sigma ^{\alpha}(\bm q) = \sum_{\bm k}c^{\dagger}_{\bm k_-}\sigma ^{\alpha} c_{\bm k_+},\\
    \hat j^{\alpha}_{\text s, i}(\bm q) = \sum_{\bm k}c^{\dagger}_{\bm k_-}j^{\alpha}_{\text s,i}(\bm k)c_{\bm k_+}, 
\end{gather}
where the Greek indices $\alpha (=x,y,z)$ specify the spin direction and the Roman indices $i(=x,y,z)$ specify the flow direction.
$j^{\alpha}_{\text s, i}=-e (\sigma ^{\alpha}v_i + \alpha _{\text R} \epsilon _{z\alpha i})$ are the vertex parts of the spin-current density operators with $v_i=k_i/m$.
The current density operators are given by the following equation:
\begin{align}
    \hat j_{\text e,i}(\bm q) = \sum_{\bm k}c^{\dagger}_{\bm k_-} j_{\text e,i}(\bm k) c_{\bm k_+},
\end{align}
where $j_{\text e,i}(\bm k)=-e(v_i+\alpha _{\text R}\epsilon _{z\alpha i}\sigma ^{\alpha})$ are the vertex parts of the current density operators.
Note that the contribution of the anomalous velocity due to $\bm u_{\bm q,\omega}$ is negligible in this study.

The expectation value in non-equilibrium states is given by the following equation:
\begin{align}
    \average{\hat j_{\text s,\mu}^{\alpha}(\bm q, \omega)}_{\text{ne}} = \int ^{\infty}_{-\infty}
    \frac{d\epsilon}{2\pi i} \sum_{\bm k} \text{tr}\left[
    j^{\alpha}_{\text s, \mu}G^{<}_{\bm k_+,\bm k_-}(\epsilon _+,\epsilon _-)
    \right],
\end{align}
where $\mu=(0,x,y,z)$ and $\epsilon _{\pm}=\epsilon \pm \frac{\omega}{2}$.
Here, the trace is taken for the spin space.
$G^<_{\bm k,\bm k'}(\epsilon, \epsilon^{\prime})$ represents the lesser component of the non-equilibrium path-ordered Green function,
$G_{\bm k,\bm k'}(t,t')=-i\average{T_{\text C}c_{\bm k}(t)c_{\bm k'}^{\dagger}(t')}_{\text{mat}}$,
where $T_{\text C}$ is a path-ordering operator, and $\average{\cdots}_{\text{mat}}
$ represents the expectation value in the non-equilibrium state

.
We expand the non-equilibrium Green function for each of the perturbation Hamiltonian $H'_{\text k}(t)$ and $H'_{\text r}(t)$ and calculate the linear response to the lattice velocity field $\bm u_{\bm q,\omega}$ up to the first order in $\bm q$ and $\omega$.

\begin{figure}[tbp]
\centering
\includegraphics[width=75mm]{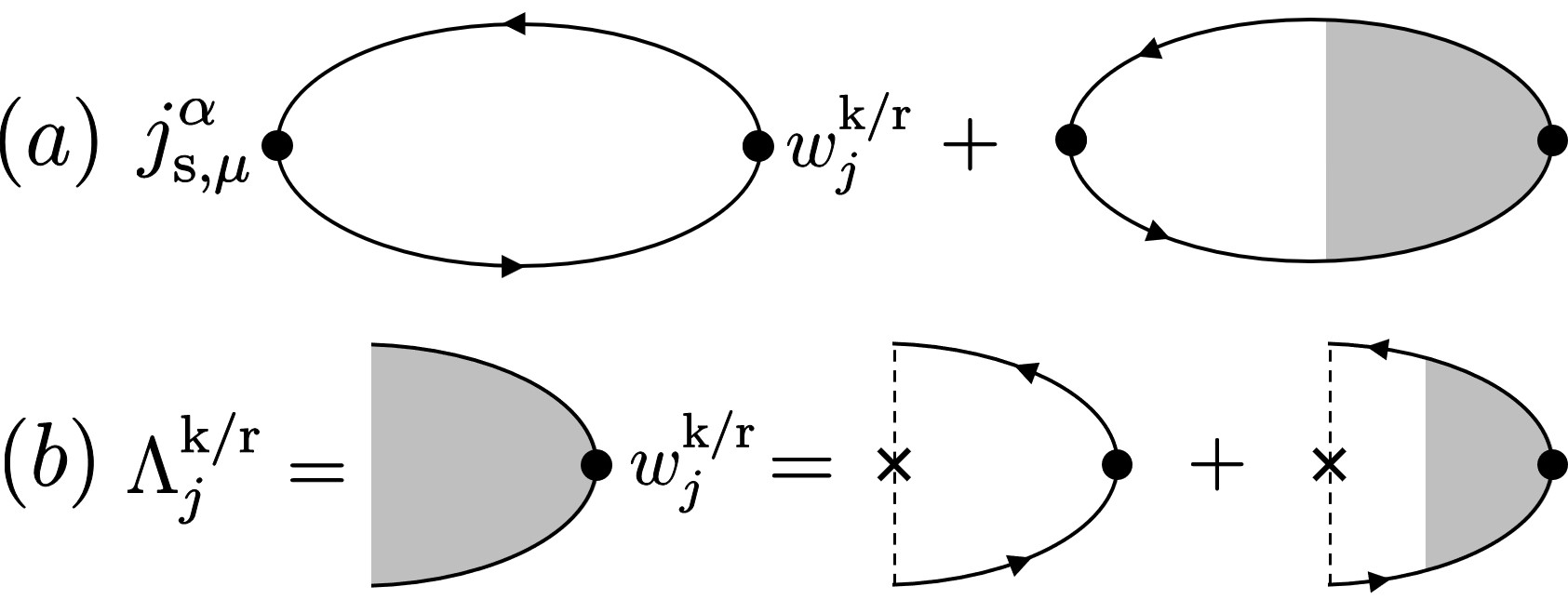}
\caption{
(a) Linear response of the spin-current density ($\mu =x,y,z$) and spin density ($\mu =0$) to the lattice velocity field $u^j_{\bm q,\omega}$.
(b) Three-point vertices with ladder vertex corrections.
The crosses and dashed lines indicate the impurities and impurity potential, respectively.
}
\label{bubble}
\end{figure}

\begin{figure}[tbp]
\centering
\includegraphics[width=50mm]{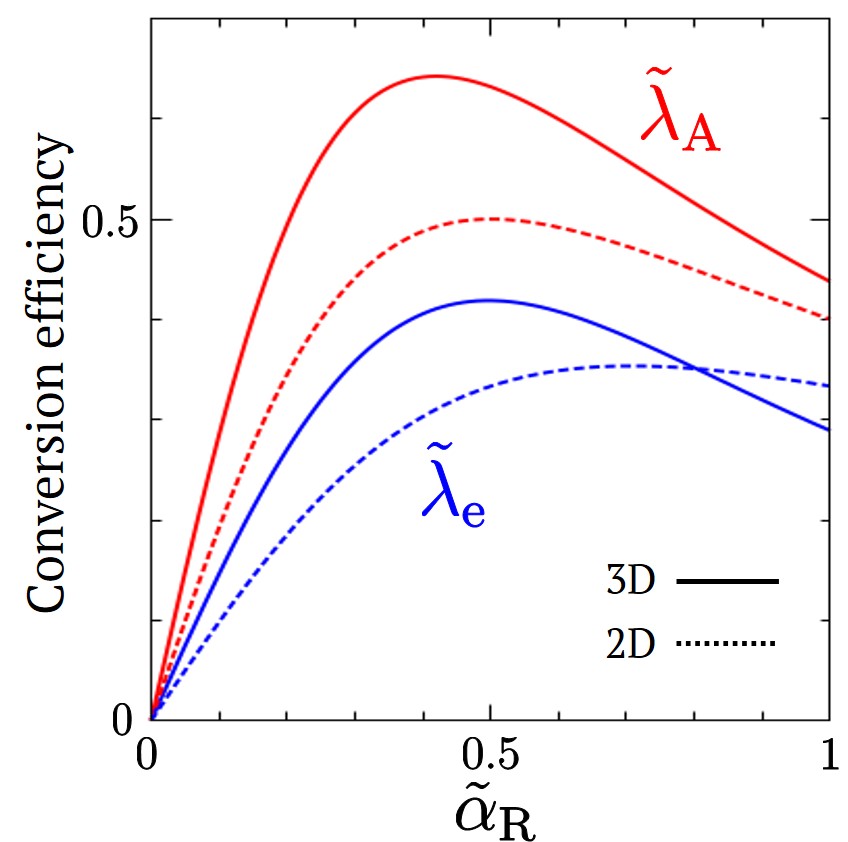}
\caption{
$\tilde \alpha _{\text R}$-dependence of the dimensionless current--spin conversion efficiency, $\tilde \lambda_{\text{A/e}} \equiv -\frac{e\mu}{k_{\text F}}\lambda_{\text{A/e}}$.
The red lines indicate the case in which the conversion occurs via mechanical means and the blue lines indicate the (electric) conventional EE.
The rigid lines represent 3D systems and the dashed lines represent 2D systems.
The results show that the conversion efficiency due to mechanical means is different from that due to the EE.
}
\label{graph1}
\end{figure}

First, we consider the so-called Fermi surface term, which can be dominant in clean metals. 
The non-equilibrium spin density and the spin-current density are given by
\begin{align}
    \average{\hat j^{\alpha}_{\text s,\mu}}^{\text{k/r}}_{\text{surf}} = \frac{i\omega}{2\pi} \sum_{\bm k} \text{tr}\left[ j^{\alpha}_{\text s,\mu}G^{\text R}_{\bm k}\tilde w^{\text{k/r}}_j G^{\text A}_{\bm k} \right] u^j_{\bm q,\omega},
\end{align}
where $\tilde w^{\text{k/r}}_j=w^{\text{k/r}}_j+\Lambda ^{\text{k/r}}_j$ with the three-point vertices $\Lambda ^{\text{k/r}}_j$ presented in Fig.~\ref{bubble}.
The spin density and spin-current density are given by the following equation:
\begin{gather}
    \average{\hat{\sigma}^{\alpha}}^{\text k}_{\text{surf}} = i\omega \alpha _{\text R} m\nu _0 \tilde a_2 \tau   [\hat z \times \bm u_{\bm q,\omega}]^{\alpha},\label{aee1}\\
    \average{\hat j^z_{\text s,i}}^{\text k}_{\text{surf}} = i\omega \frac{e\nu_0}{8}\tilde a_2 \epsilon_{zij}u^j_{\bm q,\omega},\label{she1}
\end{gather}
where $a_n = \nu _0^{-1}\sum_{s,\bm k}(s\tilde k_{\text t})^{n-1}\delta (\mu-\epsilon _{(s)})$ are dimensionless parameters with $\tilde k=k/k_{\text F}$, and $\nu_0=\frac{m}{2\pi}$ (for 2D), $\nu _0=\frac{mk_{\text F}}{2\pi ^2}$ (for 3D) are the density of states in free-electron system without the Rashba SOI with $k_{\text F}=\sqrt{2m\mu}$ being the Fermi wavenumber.
For the calculations, we also defined $\tilde a_{2n}=-\tilde \alpha _{\text R}a_{2n}$.
Note that a spin current proportional to the first order in $\bm q$ is also generated, which can be found in Ref.~\cite{helicity}.
Eqs.~(\ref{aee1}) and (\ref{she1}) show that the spin density and spin currents are generated by the acceleration of the lattice and are considered to be the AEE and the spin Hall effect, respectively.
Eq.~(\ref{aee1}) can be put in the form of $\average{\hat{\bm \sigma}}=\lambda _{\text{A}}[\hat z\times \average{\hat{\bm j}_{\text e}}]$ with the current density $\average{\hat j_{\text e,i}}$. 
Here, $\lambda _{\text{A}}$ is the current--spin conversion efficiency:
\begin{align}
    \lambda_{\text{A}} = -\tilde \alpha _{\text R}\frac{k_{\text F}}{e\mu} \left[
    \frac{a_3}{\tilde a_2} - 2\tilde \alpha _{\text R}^2 \left( 1-\frac{\tilde a_2}{2a_1} \right)
    \right],
\end{align}
where $\tilde \alpha _{\text R}=m\alpha _{\text R}/k_{\text F}$ is the dimensionless Rashba parameter.
The dependence of the conversion efficiency $\lambda _{\text A}$ on the Rashba parameter $\tilde \alpha _{\text R}$ is presented in Fig.~\ref{graph1} together with the case of the (electric) conventional EE, $\lambda _{\text e}=-\tilde \alpha _{\text R}\frac{k_{\text F}}{e\mu}(1-\frac{a_1}{\tilde a_2})[\frac{a_3}{\tilde a_2}-2\tilde \alpha_{\text R}(2-\frac{\tilde a_2}{a_1}-\frac{a_1}{\tilde a_2})]$. 
If Eq.~(\ref{aee1}) originates from the EE, $\lambda _{\text A}$ should match $\lambda _{\text e}$ because EE is due to the band shift of the systems with spin-momentum locking.
However, $\lambda _{\text A}$ and $\lambda _{\text e}$ show different $\alpha _{\text R}$-dependencies, which indicates that the AEE is a different mechanism from the conventional EE due to the electric field.
The current density generated by the lattice distortion is given by the following equation:
\begin{gather}
    \average{\hat j_{\text e,i}} = -i\omega e\mu \nu _0 \tau
    \bigg\{
    2c_1 \delta ^z_{ij} \nonumber  \\
    +\biggl[  a_3 - 2 \tilde \alpha _{\text R} ^2 \tilde a_2\left( 1 - \frac{\tilde a_2}{2a_1} \right) \biggr]
     \delta ^{2\text d}_{ij}
    \biggr\} u_{\bm q,\omega}^j,
\end{gather}
where $\delta ^z_{ij}=\delta _{iz}\delta _{jz}$, $\delta ^{\text{2d}}_{ij}=\delta _{ix}\delta _{jx}+\delta _{iy}\delta _{jy}$, and $c_n = \nu _0\sum_{s,\bm k}\tilde k_z^2 (s\tilde k_{\text t})^{n-1}\delta (\mu -\epsilon_{(s)})$.
The term independent of $\alpha _{\text R}$ in the current density is consistent with the results obtained for the system without the Rashba SOI: $\average{\hat{\bm j}_{\text e}}=-i\omega en_{\text e0}\tau \bm u_{\bm q,\omega}$ for the $\alpha _{\text R}=0$ limit with $n_{\text e0}=\frac{2k_{\text F}^2}{3m}\nu_0$ being the electron number density for systems without the Rashba SOI. 

It is well-known that the spin Hall effect due to the uniform and static external field vanishes\cite{rashba2003,dimitrova2005,shi2006}.
Conversely, the present spin Hall currents remain as presented in Eq.~(\ref{she1}), since it is caused by the lattice acceleration associated with the surface acoustic waves. 

\begin{figure}[tbp]
\centering
\includegraphics[width=50mm]{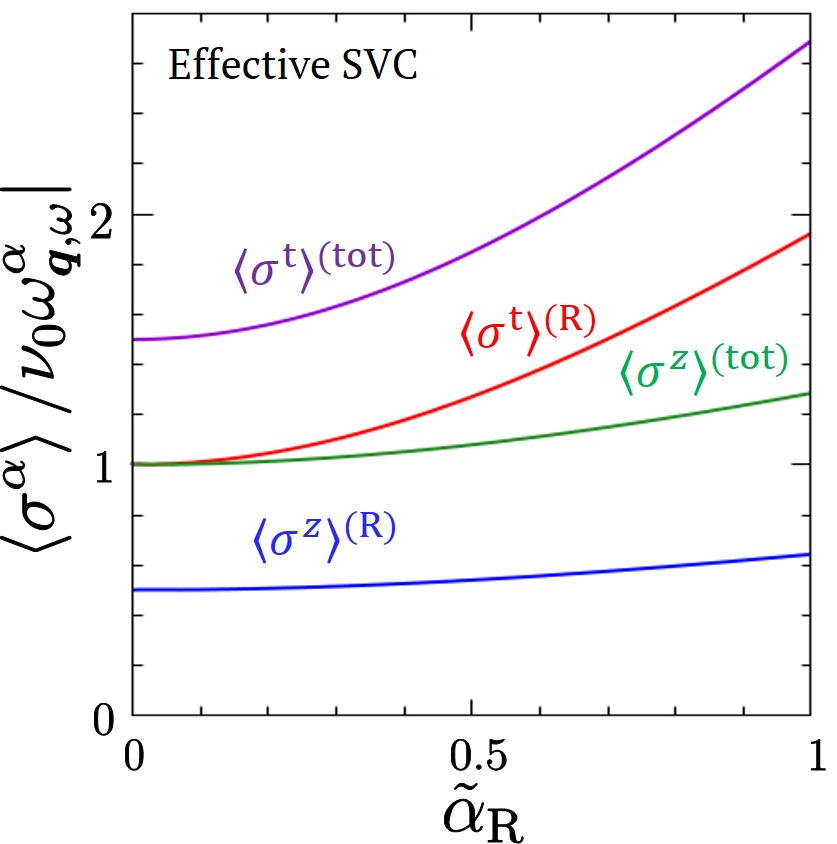}
\caption{
$\tilde \alpha _{\text R}$-dependence of the dimensionless current-spin conversion efficiency, $\tilde \lambda_{\text{A/e}} \equiv \frac{\mu}{k_{\text F}}\lambda_{\text{A/e}}$.
The red lines represent the case of mechanical means and the blue lines represent (electric) conventional EE.
The rigid lines represent 3D systems and the dashed lines represent 2D systems.
The results show that the conversion efficiency due to mechanical means is different from that due to EE.
}
\label{graph2}
\end{figure}

Next, we consider the so-called Fermi sea term, which contributes only to the spin density response.
\begin{gather}
    \average{\hat \sigma ^{\alpha}}^{\text r}_{\text{sea}} = \frac{i\omega}{4\pi} \sum_{\bm k} \int ^0_{-\infty} d\epsilon \text{tr}
    \biggl[ \sigma ^{\alpha} \partial _{\epsilon}G^{\text R}_{\bm k}(\epsilon ) w^{\text r}_j G^{\text R}_{\bm k} (\epsilon) \nonumber \\
    - \sigma^{\alpha}G^{\text R}_{\bm k}(\epsilon ) w^{\text r}_j \partial _{\epsilon} G^{\text R}_{\bm k} (\epsilon)
    - \average{\text R\leftrightarrow \text A}\biggr] u^j_{\bm q,\omega}.
\end{gather}
This term is calculated using the following equation:
\begin{gather}
    \average{\hat \sigma ^{\alpha}} ^{\text{r}}_{\text{sea}}= -\frac{\nu}{4} \biggl[
     \left( 1-\frac{\tilde a_2}{a_1} \right) \omega _{\bm q, \omega}^z\delta _{\alpha z}
     \nonumber \\
    - \left(
    \frac{15}{4}-\frac{7\tilde a_2}{8a_1} - \frac{\tilde \alpha _{\text R}^2\mu }{4\nu} \frac{d}{d\mu} \tilde a_2 \nu_0
    \right)  \omega _{\bm q, \omega}^{\alpha} \delta _{\alpha \neq z} \biggr],\label{svc1}
\end{gather}
where $\bm \omega _{\bm q, \omega}= i\bm q\times \bm u_{\bm q,\omega}$ is the vorticity of the lattice distortion.
Eq.~(\ref{svc1}) suggests that the spin density is induced by the vorticity motion of the lattice via the Rashba SOI, which can be referred to as the RSVC.
This includes the terms that do not depend on the Rashba SOI, i.e. $\average{\hat \sigma ^{\alpha}}^{\text r}_{\text{sea}}=\nu _0(\frac{1}{2}\omega ^z_{\bm q,\omega} + \omega _{\bm q,\omega}^{\alpha}\delta_{\alpha \neq z})$ for the non-Rahsba limit $\alpha _{\text R}\rightarrow 0$,
 however, the RSVC requires the Rashba SOI.
The dependence of the spin density induced via the RSVC on the Rashba parameter is presented in Fig.~\ref{graph2} for the 3D system.
Here, $\average{\hat \sigma ^{\alpha}}^{\text{tot}}=\average{\hat \sigma ^{\alpha}}^{\text r}+\average{\hat \sigma ^{\alpha}}^{\text{sv}}$ is the total spin density due to the RSVC and the conventional SVC.
As the RSVC does not occur in the absence of the Rashba SOI, the RSVC can be attributed to the spatial symmetry breaking due to the Rashba SOI.
The spin density in response to the SVC $H_{\text{sv}}=-\frac{1}{4}\bm \sigma \cdot \omega _{\bm q,\omega}$ is given by $\average{\hat \sigma ^{\alpha}}^{\text{sv}}=\frac{1}{4}\chi ^{\text s}_{\alpha \beta}\omega ^{\beta}$, where $\chi ^{\text s}_{\alpha \beta}=\frac{\nu_0}{2}[\delta ^{\text{2d}}_{\alpha \beta}\tilde a_2+2\delta ^z_{\alpha \beta}(\tilde a_2-a_1)]$ is the spin-spin correlation function.
Note that the spin-spin correlation function approach to the Pauli paramagnetism is $\chi ^{\text s}_{\alpha \beta}=2\nu _0\delta _{\alpha \beta}$ for $\alpha _{\text R}\rightarrow 0$.
In the 2D system, the spin density is constant and independent of the Rashba parameter.
As presented in Eq.~\ref{svc1}, the spin density induced by the RSVC is comparable to or larger than that caused by the conventional SVC.
Therefore, it is expected that the combination of the RSVC and the conventional SVC provides more efficient spin generation in the Rashba systems.

Finally, we show that Eq.~(\ref{svc1}) can be expressed in terms of the Berry curvature-like quantity according to the following equation:
\begin{gather}
    \average{\sigma ^{\alpha}}_{\text{sea}}^{\text r} = \frac{i\omega}{4} \sum_{\bm k,s} f(\epsilon _s) \Omega ^{\alpha}_{s,j}(\bm k) u^j_{\bm q,\omega},
\end{gather}
where $f(\epsilon _s)$ is the Fermi distribution function.
$\Omega ^{\alpha}_{s,j}(\bm k)$ is similar to the Berry curvature:
\begin{align}
    \Omega ^{\alpha}_{s,j}(\bm k)=\frac{2\text{Im}[\sigma ^{\alpha}_{s\bar s}w ^{\text r}_{j,\bar ss}]}{(\epsilon _s -\epsilon _{\bar s})^2},
\end{align}
where $\bar s=-s$.
Here, $\sigma ^{\alpha}_{s\bar s}$ and $w^{\text r}_{j,s\bar s}$ are the matrix elements of the spin density and the vertex part given in Eq.~(\ref{wr}) for the eigenstates of $\epsilon _s$ and $\epsilon _{\bar s}$, respectively.

In summary, we theoretically studied the generation of spin density from the lattice distortion dynamics in 2D and 3D Rashba systems. 
The results indicate that there are two mechanisms for the spin density induction: one is the AEE due to the lattice acceleration and the other is the RSVC due to the lattice vorticity.
It is suggested that the AEE allows for more efficient spin-charge conversion than the conventional EE.
The spin density due to the RSVC remained nonzero even in the non-Rashba limit $\alpha _{\text R}\rightarrow 0$.
This induced spin dnsity could be attributed to the spatial symmetry breaking due to the Rashba SOI and was expressed in terms of the Berry curvature-like quantity.
The spin density induced by the RSVC is suggested to be comparable to or larger than that induced by the conventional SVC.
Since the RSVC and the conventional SVC are cooperative, a more efficient spin generation is envisaged in the Rashba systems.

We would like to thank Y. Nozaki, K. Yamanoi, and T. Horaguchi for enlightening discussions.
We also thank H. Kohno, A. Yamakage, Y. Imai, J. J. Nakane. Y. Yamazaki, Y. Ogawa, Y. Ozu, and Y. Hayakawa for daily discussions.
This work was partially supported by JST CREST Grant Number JPMJCR19J4, Japan.
TF is supported by Grant-in Aid for JSPS Fellows Grant Number 19J15369, and by a Program for Leading Graduate Schools "Integrative Graduate Education and Research in Green Natural Sciences".
MM is supported by JSPS KAKENHI for Grants (Nos. 20H01863 and 21H04565) and the Priority Program of the Chinese Academy of Sciences, Grant No. XDB28000000.

\end{document}